\documentclass[a4paper]{jpconf}
\usepackage{graphicx}
\usepackage{iopams}

\begin{document}
\title{Tidally-induced thermonuclear Supernovae}

\author{Stephan Rosswog$^1$, Enrico Ramirez-Ruiz$^2$, W. Raphael Hix$^3$}

\address{$^1$ School of Engineering \& Science, Jacobs University Bremen, 28759 Bremen, Germany\\
                $^{2}$ Dept. of Astronomy \& Astrophysics, University of California, Santa Cruz, CA 95064, USA\\ 
                $^{3}$ Physics Division, Oak Ridge National Laboratory, Oak Ridge, TN37831-6374, USA}

\ead{srosswog@jacobs-university.de}

\def\paren#1{\left( #1 \right)}
\def\Mesz{M\'esz\'aros~}
\def\Pacz{Paczy\'nski~}
\def\Kluz{Klu\'zniak~}
\def\p{$e^\pm \;$}
\def\msun{M$_{\odot}$}
\def\Msun{M$_{\odot}$ }
\def\be{\begin{equation}}
\def\ee{\end{equation}}
\def\bi{\begin{itemize}}
\def\ei{\end{itemize}}
\def\bea{\begin{eqnarray}}
\def\eea{\end{eqnarray}}
\def\gcc{gcm$^{-3}$ }
\def\sun{\odot}

\begin{abstract}
We discuss the results of 3D simulations of tidal disruptions of white dwarfs by moderate-mass black holes as  they 
may exist in the cores of globular clusters or dwarf galaxies. Our simulations follow self-consistently the hydrodynamic and
nuclear evolution from the initial parabolic orbit over the disruption  to  the build-up of an accretion disk around the black hole.  
For strong enough encounters (pericentre distances smaller than  about 1/3 of the tidal radius) the tidal compression is 
reversed by a shock and finally results in a thermonuclear explosion. These explosions are not restricted to progenitor 
masses close to the Chandrasekhar limit, we find exploding examples throughout the whole white dwarf mass range. 
There is, however, a restriction on the masses of the involved black holes: black holes more massive than 
$2\times 10^5$ \Msun swallow a typical 0.6 \Msun white dwarf before their tidal forces can overwhelm the star's self-gravity. 
Therefore, this mechanism is characteristic for black holes of  moderate masses. The material that remains bound to the black 
hole settles into an accretion disk and produces an X-ray flare close to the Eddington limit of $L_{\rm Edd} \simeq 10^{41} {\rm erg/s}\;
(M_{\rm bh}/1000 M_\sun)$, typically lasting for a few months. The combination of a peculiar thermonuclear supernova 
together with an X-ray flare thus whistle-blows the  existence of such moderate-mass black holes. The next generation of wide 
field  space-based instruments should be able to detect such events.
\end{abstract}

\section{Introduction}
Two classes of black holes are established beyond reasonable doubt: black holes of a few solar masses ("stellar-mass") 
 make themselves known by interacting with a companion star \cite{mcclintock03} and "supermassive" black holes with masses
of $10^6 - 10^{10}$ \Msun seem to  lurk in the centres of most, if not all galaxies, see e.g. \cite{kormendy95,ferrarese00}. A third class, 
so-called "intermediate-mass black holes"  (IMBH) with masses of $\sim1000$ \msun, represents a plausible, yet to date still missing 
link. The evidence for their existence is mounting, but  to date it is  not entirely conclusive.  Based on kinematical studies Gerssen et al. 
\cite{gerssen02,gerssen03} have suggested the presence of a $3.9 \times 10^3$ \Msun black hole in the core of the globular cluster M15. 
More recently, there were further reports on IMBHs in the globular clusters NGC 2808 \cite{maccarone08} and omegaCen \cite{noyola08}.
Additional evidence comes from  ultraluminous, compact X-ray sources in young star clusters \cite{zezas02,pooley06} and from n-body 
simulations \cite{portegies04} that indicate that runaway collisions in dense young star clusters can produce rapidly growing
massive stars that should ultimately collapse to form IMBHs.
None of the above arguments is strong enough to close the case, but their different nature gives this hypothesis credibility.\\
At the very large number densities in the environments that may habor such black holes  stars are disrupted at a rate of 
$\sim 10^{-7} {\rm yr}^{-1}\; M_{\rm bh, 3}^{4/3}  \; n_{\ast,6}  \; \sigma_{10}^{-1}   \;  \left( R_{\rm per}/R_{\rm t}\right)$, where $M_{\rm bh, 3}$ 
is the black hole mass in units of 1000 \msun, $n_{\ast,6}$ is the central stellar density in units of $10^6 \; {\rm pc^{-3}}$, 
$\sigma_{10}$ the velocity dispersion in $10 \; {\rm km s^{-1}}$ and $R_{\rm per}$ is the distance of closest approach. The quantity 
$R_{\rm t}= \left( M_{\rm bh}/M_{\ast}\right)^{1/3} R_{\ast}$ is the tidal radius, the approximate separation where the hole's
tidal field wins the battle against the star's self-gravity. It is proportional to the stellar radius $R_{\ast}$, but only grows $\propto M_{\rm bh}^{1/3}$,
i.e. substantially slower than the gravitational radius, $R_{\rm S}= 2 G M _{\rm bh}/c^2$, of the black hole. As a consequence, too massive
black holes, $ M _{\rm bh} > M_{\rm bh, lim} \simeq 2.5 \times 10^{5} M_\odot \; R_{\rm wd,9}^{3/2}
M_{\rm wd, 0.6}^{-1/2}$ will swallow stars before disrupting them. Here, $R_{\rm wd,9}$ is the white dwarf radius in $10^9$ cm and
$M_{\rm wd, 0.6}$ its mass in units of 0.6 \msun. Thus, white dwarfs with their small radii and 
their (at least in principle) available nuclear fuel can only be tidally squeezed and disrupted by black holes of moderate masses.
Therefore they are precious tools to probe the existence of this type of black hole. \\
Luminet and Pichon were the first to discuss the tidal disruption of white dwarfs by black holes \cite{luminet89a} in the context of the
affine star model \cite{carter83,carter85}.  Wilson and Mathews \cite{wilson04} believe that general relativistic effects may lead to a 
tidal ignition even for supermassive black holes and at separations of many gravitational radii. A hydrodynamic simulation with a 
Newtonian Adaptive Lagrangian Eulerian (ALE) code was performed recently \cite{dearborn05} where the gravitational constant was 
adjusted to mimic general relativistic effects. We have recently performed a large set of 3D hydrodynamic simulations 
\cite{rosswog08a,rosswog08b,rosswog08d} that incorporate the nuclear energy generation self-consistently with the hydrodynamics.

\section{Ingredients for the simulations}
Here we briefly summarize the ingredients of our simulations, both in terms of physics and 
numerical methods. We described the numerical aspects in 
 \cite{rosswog08b} and in \cite{rosswog08d} we concentrated on the 
 physics and the astrophysical implications of such tidal disruptions.\\
We use the smoothed particle hydrodynamics method (SPH) to follow the hydrodynamic 
evolution of the white dwarf matter.  Being completely Lagrangian and conserving mass, energy, momentum 
and angular momentum by construction the method is ideal for the study of such a violent stellar disruption. 
General reviews on the method can be found in \cite{benz90a,monaghan92,monaghan05}, 
our particular implementation is documented in \cite{rosswog08b}.
We calculate the forces from self-gravity via a binary tree \cite{benz90b} and those from the central black hole
via a relativistic pseudo potential \cite{paczynski80}, details of how the singularity is treated numerically are laid out
in \cite{rosswog05a}.%
We close the system of hydrodynamics equations via the HELMHOLTZ equation of state (EOS) \cite{timmes00a}. It 
accepts an externally calculated chemical composition, facilitating the coupling to nuclear 
reaction networks. The electron-/positron equation of state makes no assumptions about the degree 
of degeneracy or relativity, the exact expressions are integrated numerically and are subsequently tabulated.
The interpolation in this table enforces thermodynamic consistency by construction \cite{timmes00a}. 
The nuclei in the gas are treated as a Maxwell-Boltzmann gas, the photons as blackbody radiation.
Most importantly in this context, we couple a minimal nuclear reaction network \cite{hix98} to the 
hydrodyamics. Throughout this simulation set we assume a uniform nuclear composition accross the 
white dwarfs. Stars with masses $<$ 0.6 \Msun are instantiated as pure helium, more massive stars are 
modeled with a mass fraction of 50\% carbon and  50\% oxygen. All simulations are started from parabolic 
orbits, the strength of an encounter is parametrized by the penetration factor $\beta=R_{\rm t}/R_{\rm per}$. 
For more details we refer the interested reader to \cite{rosswog08b,rosswog08d}.

\section{Results}

\subsection{Explosion mechanism}
\begin{figure}[htbp] %  figure placement: here, top, bottom, or page
   \centering
  \centerline{\includegraphics[width=4.2in]{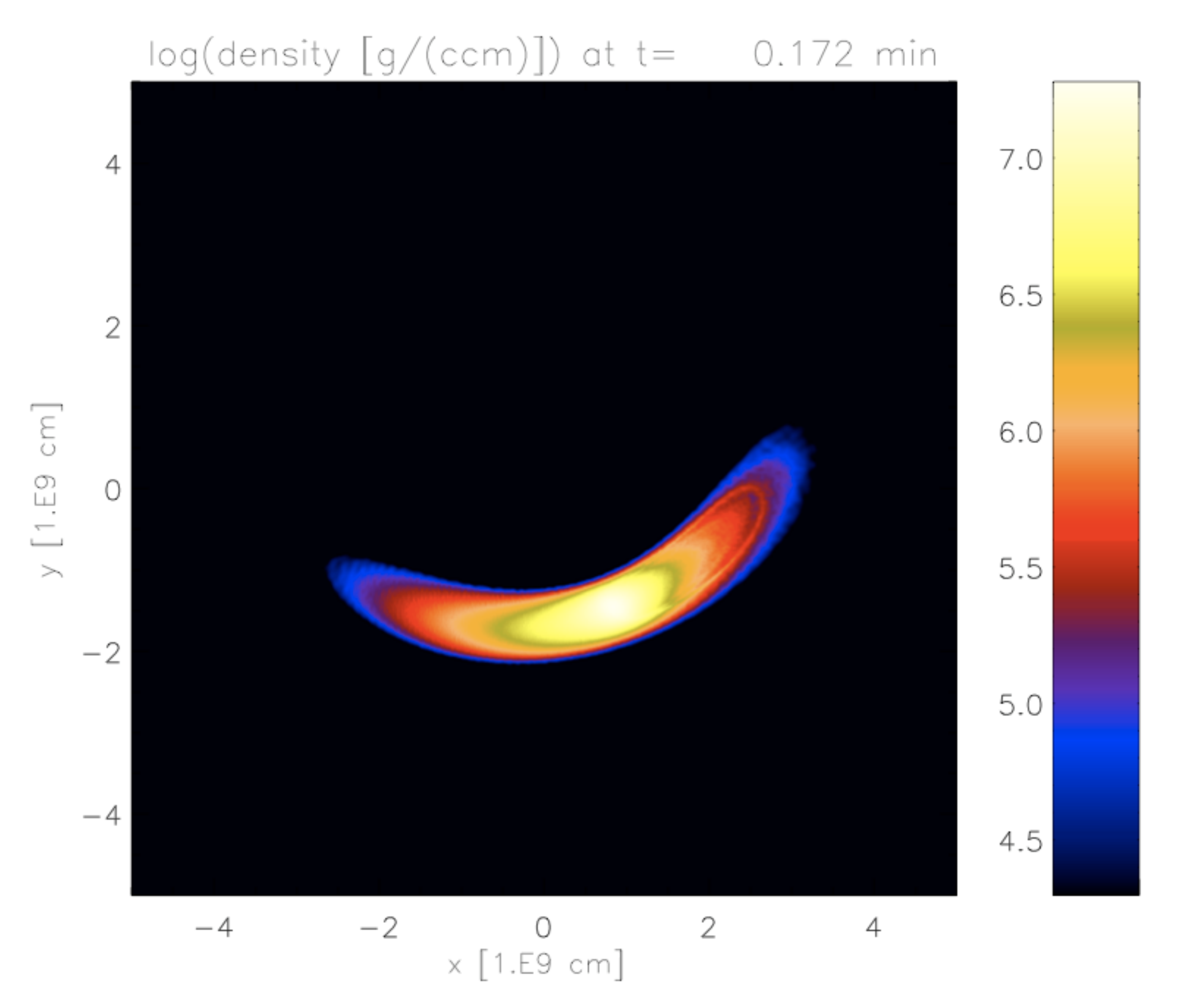} \hspace*{-0.5cm}\includegraphics[width=4.3in]{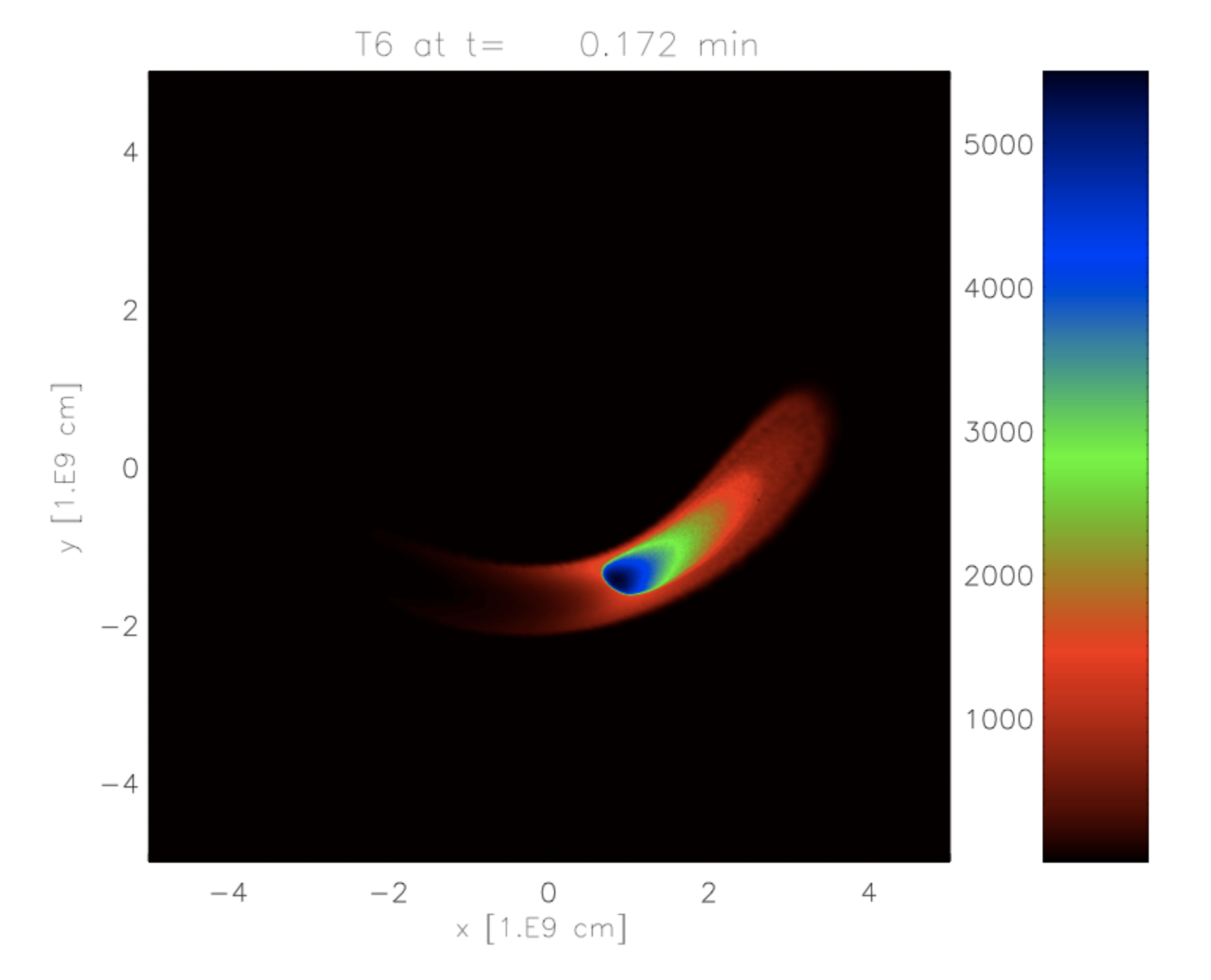} }
  \centerline{\includegraphics[width=4.2in]{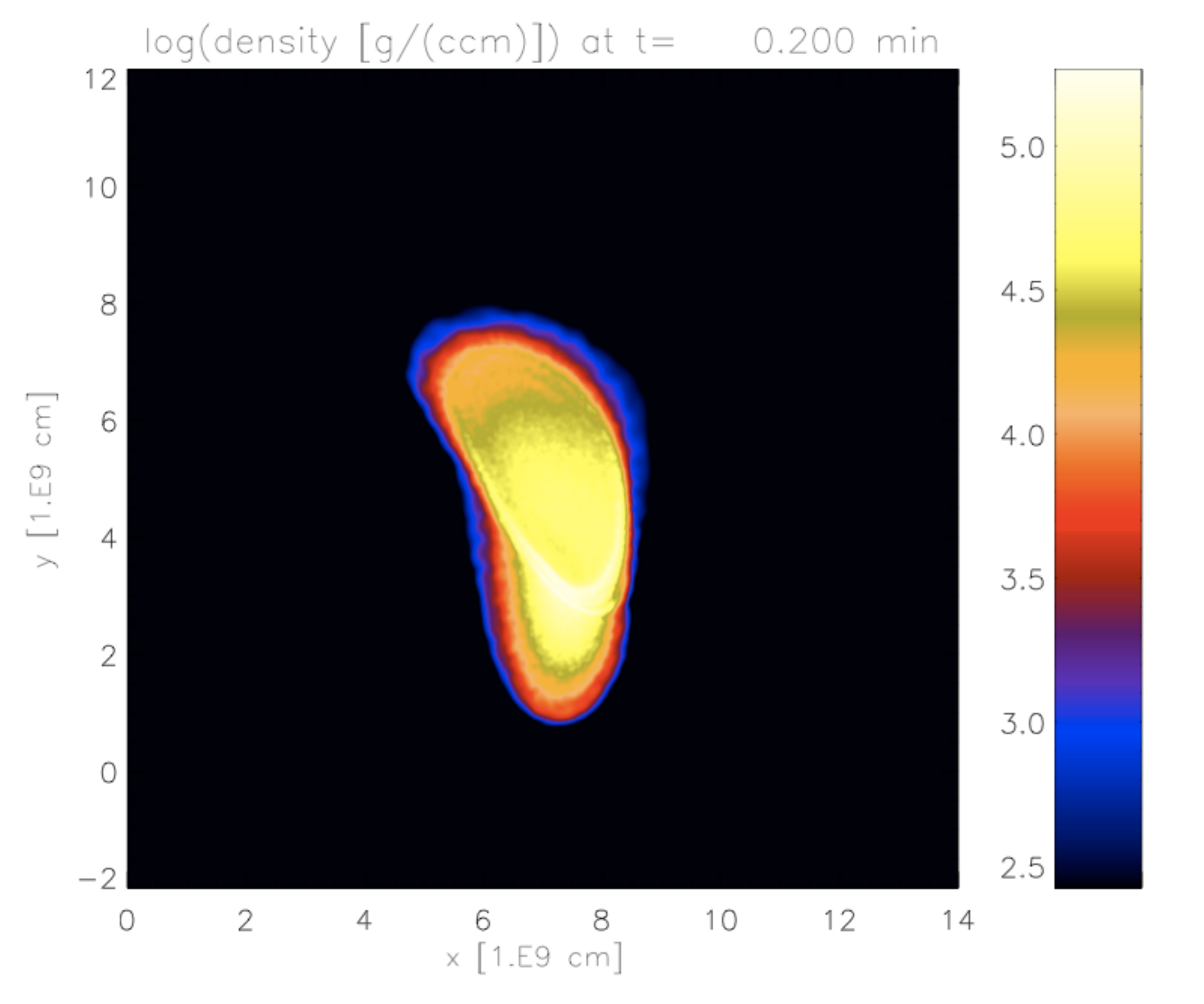} \hspace*{-0.5cm}\includegraphics[width=4.2in]{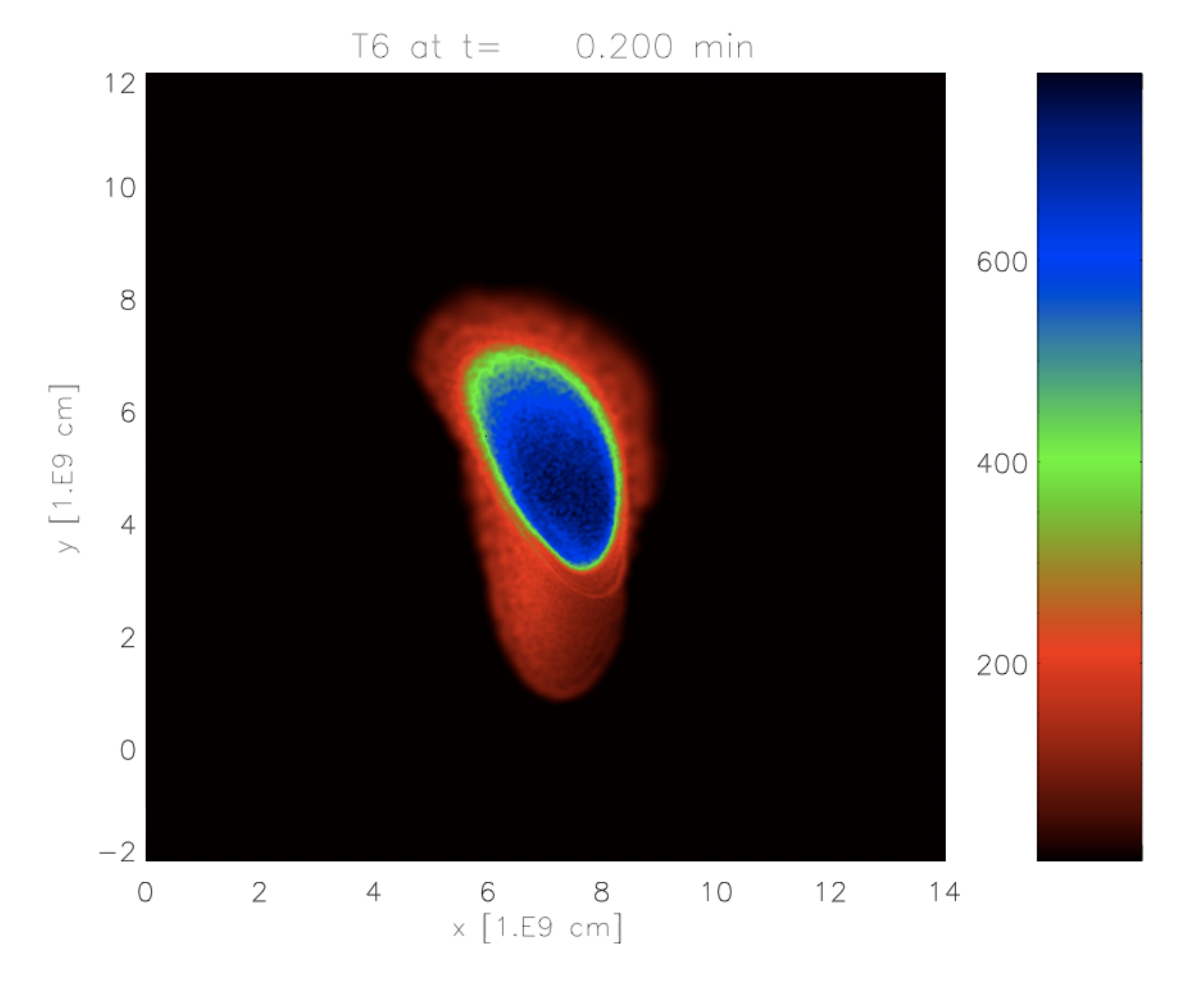} }      
   \caption{Density (left column) and temperature (right column) evolution during the disruption of 0.6 \Msun CO white dwarf by a 
   500 \Msun black hole. The distance of closest approach was 1/5 of the tidal radius.  }
   \label{fig:rho_T}
\end{figure}

The tidal gravitational field is, to lowest order, determined by the second-order tidal tensor and the third-order deviation tensor. The latter
determines the deviation of an extended body from a point particle trajectory of the same mass.
The tidal tensor determines how the fluid will be shaped. It possesses one positive and two negative
eigenvalues, see e.g. \cite{brassart08}, and, as a consequence, the tidal field stretches the star in one eigen-direction and compresses
it along the other two. Most severely, the star is compressed perpendicular to the orbital plane, this process halts when a shock forms 
\cite{kobayashi04a,brassart08,rosswog08d} that reverts the compression into an expansion. For deep enough penetrations matter
reaches nuclear statistical equilibrium, see Fig.~\ref{fig:rho_T}, upper right panel. The star is squeezed 
through a point of maximum compression on a time scale of $\tau_{\rm comp} \sim {R_{\rm wd}/v_{\rm p}} \simeq 0.2 {\rm s}\;
\left(M_{\rm wd, 0.6}\right)^{-1/6} \left(R_{\rm wd, 9}\right)^{3/2} \left(M_{\rm bh, 3} \right)^{-1/3},$ where 
$ v_{\rm p}\sim (R_{\rm g}/R_{\tau})^{1/2} c   \simeq 5 \times 10^{9} {\rm cm\;s^{-1}} \; M_{\rm wd, 0.6}^{1/6} \; R_{\rm wd, 9} ^{-1/2} 
    M_{\rm bh, 3}^{1/3} $ is the peri-centre passage velocity. This velocity exceeds 
thermonuclear flame speeds by orders of  magnitude, therefore, flame propagation effects can be safely neglected. The compression 
time scale needs to be compared to the dynamical time scale of the star, $\tau_{\rm dyn}= (G\bar{\rho})^{-1/2}\simeq 7.2 \; {\rm s} \; 
M_{\rm wd, 0.6}^{-1/2} \; R_{\rm wd, 9} ^{3/2} $, with $\bar{\rho}$ being the average stellar density and to the nuclear reaction time scale, 
$\tau_{\rm nuc}$. Only for $\tau_{\rm nuc} \ll \tau_{\rm comp}, \tau_{\rm dyn} $ can a substantial nuclear energy release be expected. 
Our large simulation set \cite{rosswog08d} shows that white dwarfs of all masses can be thermonuclearly exploded provided that the 
penetration factor exceeds $\beta \simeq 3$. For definiteness, we illustrate here the results at the example of a typical 0.6 \Msun 
carbon-oxygen (CO) white dwarf that passes a 500 \Msun black hole with a penetration factor of  $\beta=5$, see Fig. ~\ref{fig:rho_T}.

\subsection{Nucleosynthesis}
\begin{figure}[htbp] %  figure placement: here, top, bottom, or page
   \centering
  \centerline{\includegraphics[width=4.2in]{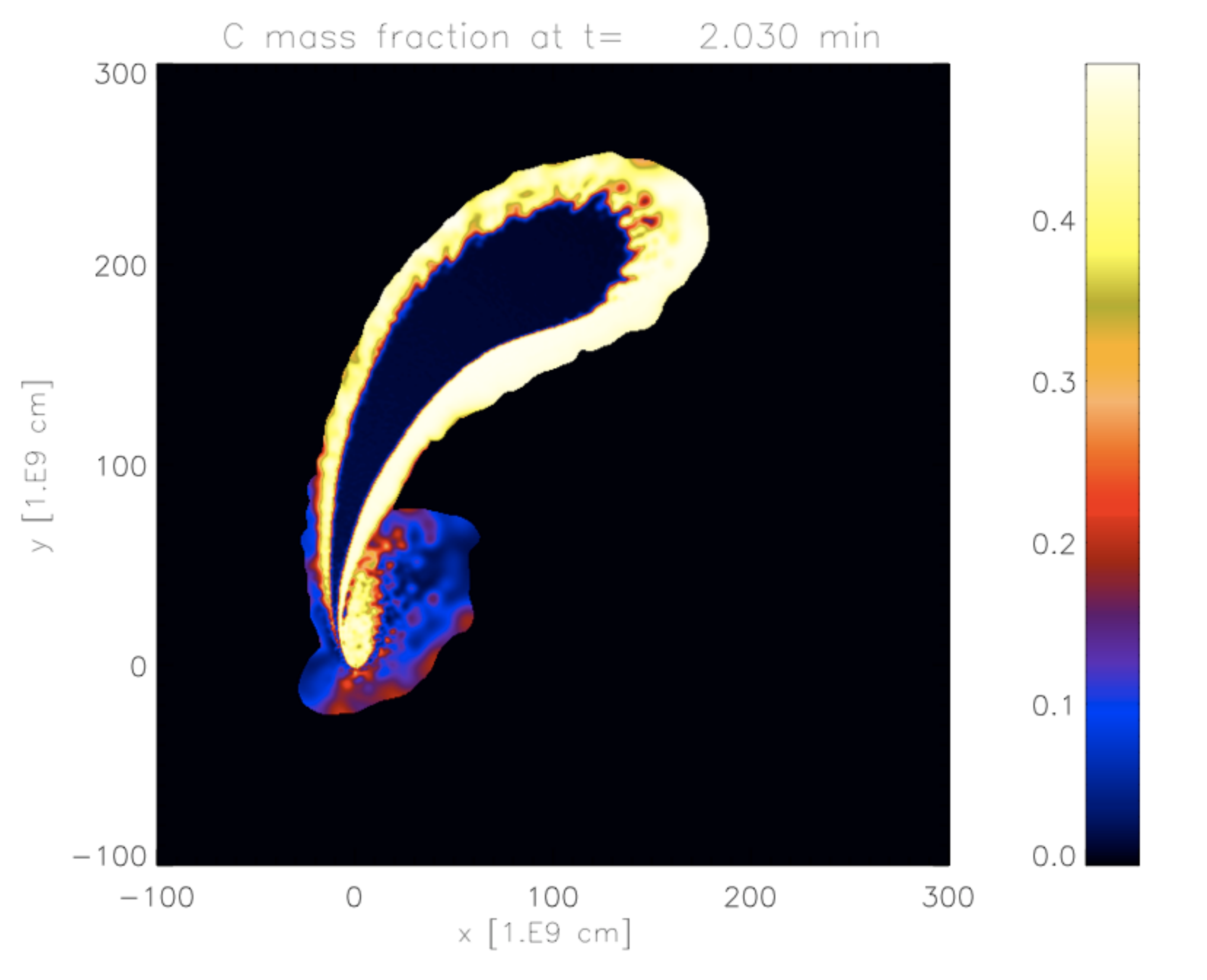} \hspace*{-0.5cm}\includegraphics[width=4.2in]{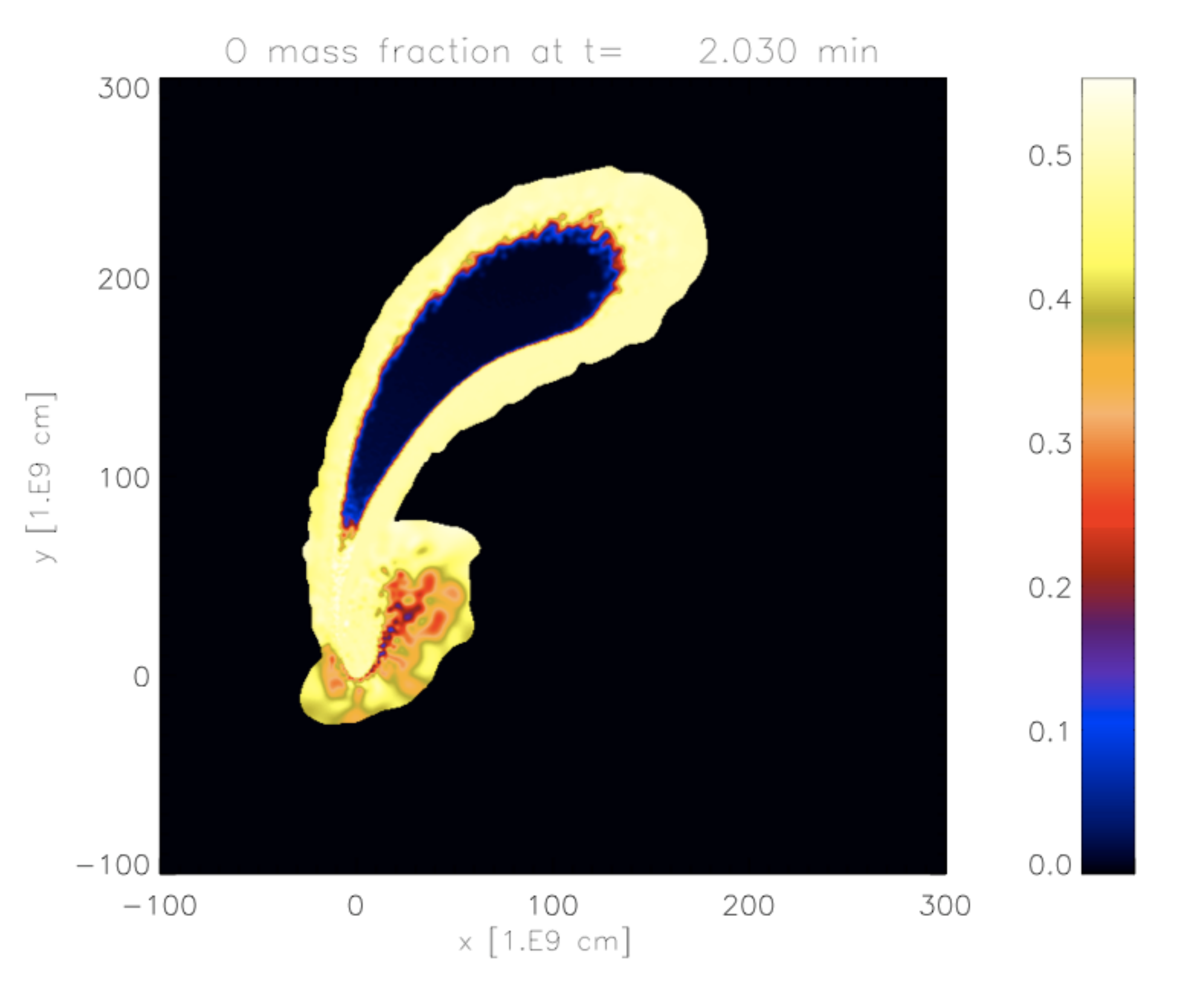} }
  \centerline{\includegraphics[width=4.2in]{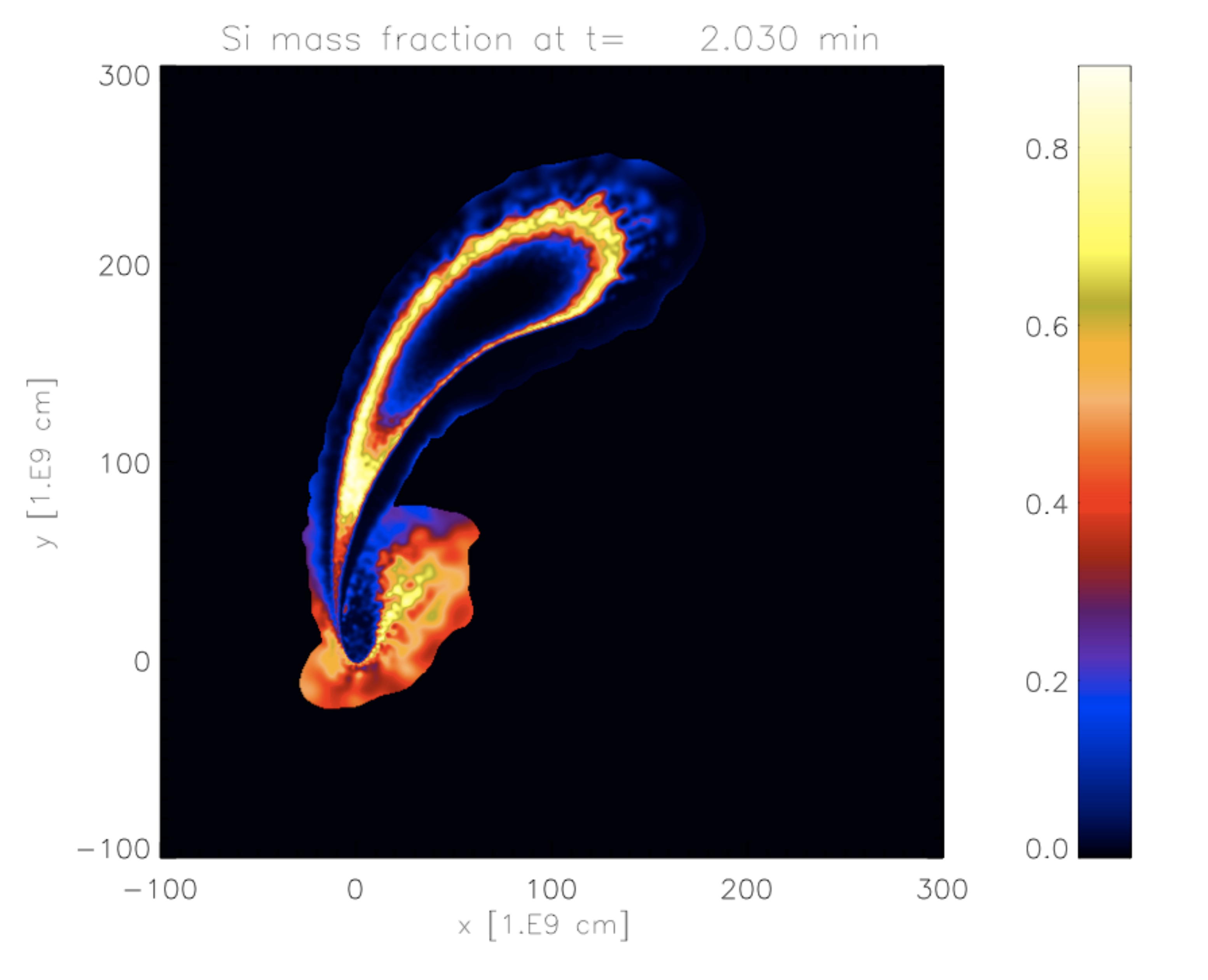} \hspace*{-0.5cm}\includegraphics[width=4.2in]{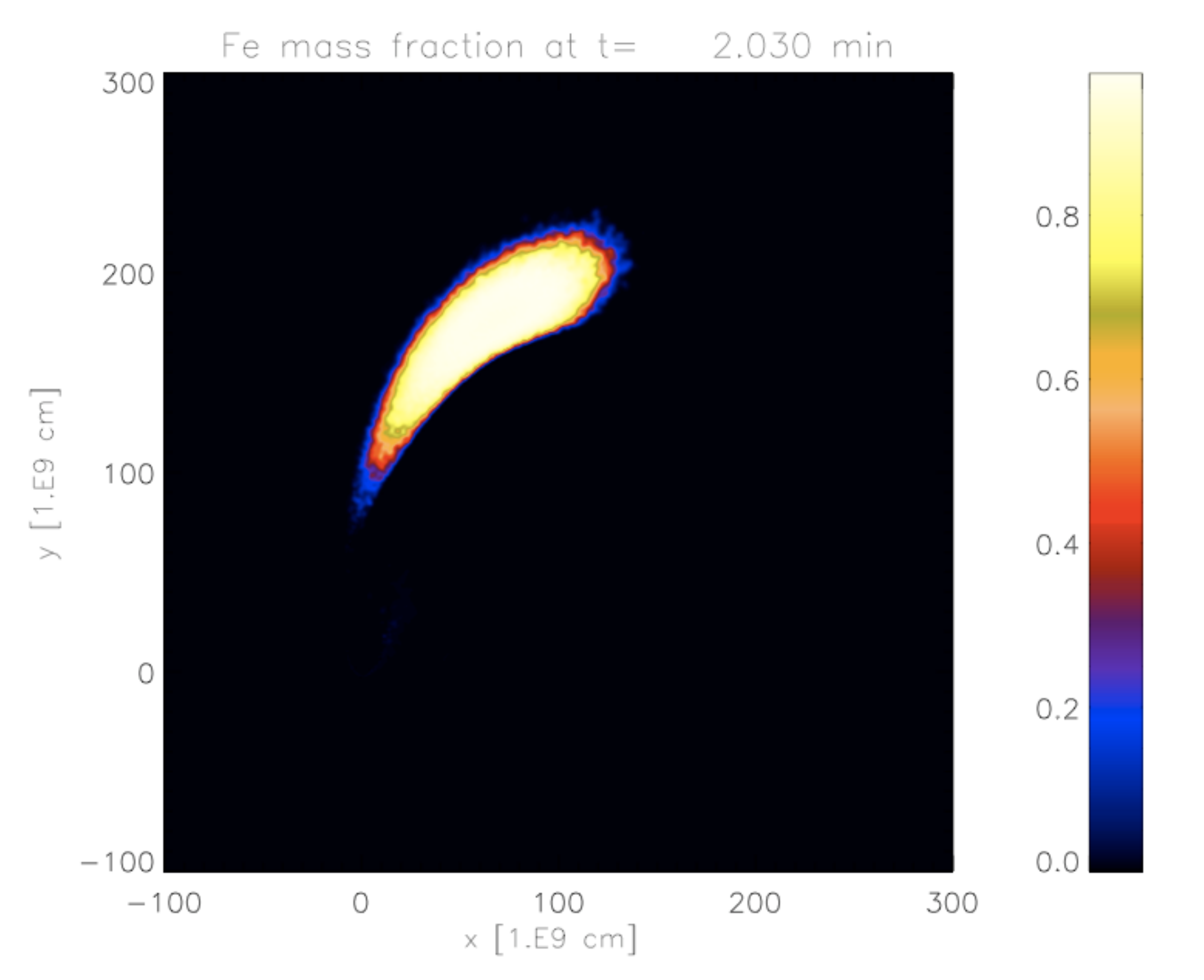} }      
   \caption{Mass fractions resulting from the disruption of 0.6 \Msun CO white dwarf by a 500 \Msun black hole. The distance of closest 
   approach was 1/5 of the tidal radius. }
   \label{fig:nucl}
\end{figure}
Nucleosynthesis is triggered predominantly in the point of maximum compression, see  Fig. ~\ref{fig:rho_T}, upper right panel, the 
concomitant nuclear energy release inflates a hot bubble in the debris centre, see Fig.~\ref{fig:rho_T}, lower panels. The 
high-density centre of the star produces an iron-group core of 0.18 \msun \footnote{Note that our network uses element 
groups, individual isotopes would need to be recovered in a post-processing step with  larger network.}, if  mainly composed of $^{56}$Ni 
this amount is comparable, but at the lower end of what is deduced for standard type Ia supernovae \cite{mazzali07}. The iron core is
surrounded by a 0.21 \Msun shell of silicon-group elements which in turn is covered by a sheath of unburned carbon-oxygen material, see 
Fig.~\ref{fig:nucl}. In reality, such a disruption would result in even stronger carbon-enhanced outer layers than shown in Fig.~\ref{fig:nucl}. 
For simplicity, we have instantiated our initial carbon oxygen white dwarf models  (M$_{\rm WD} \ge 0.6$ \msun)  as homogeneously mixed 
stars with a 50\% mass fraction of each nucleus. While such internal chemical profiles are likely accurately realized in nature in very massive white dwarfs 
($\sim 1$ \msun) \cite{mazzitelli86}, for lower masses the gravothermal adjustment of the interior during the cooling phase produces 
oxygen-enhanced stellar cores surrounded by very carbon-rich mantles ($X_C \sim 0.8$). The exact radial 
distribution depends on the exact value of $^{12}C(\alpha,\gamma)^{16}O$ rate and the details of how convection proceeds 
\cite{mazzitelli86,salaris97,straniero03}, but this general stratification tendency is  well-established. Thus, the disruption of a standard 
0.6 \Msun white dwarf should produce a highly carbon-enriched remnant atmosphere.\\
We found exploding examples throughout the whole white dwarf mass range. An extreme case is a low-mass white dwarf  that can 
(due to its large stellar and correspondingly large tidal radius) be sent very deeply into the tidal radius of a black hole. A 0.2 \msun, pure 
helium white dwarf disrupted by a 1000 \Msun black hole ($\beta=12$) produced an explosion with about 0.03 \Msun of iron-group elements. 
At the other extreme,  a 1.2 \Msun CO white dwarf exploded with 0.66 \Msun of iron group elements after passing a 500 \Msun black hole with 
$\beta=2.6$. For more details and examples, the interested reader should consult \cite{rosswog08d}. A detailed comparison of the lightcurves of 
tidally-induced supernovae and "normal" type Ia supernovae is currently being prepared \cite{kasen08}.

\subsection{Accompanying signatures}
If the release of nuclear binding energy is disregarded, about half the stellar mass is ejected, the 
rest is still bound to the black hole and, in principle, available to be accreted. Nuclear energy 
release during pericentre passage can substantially increase the amount of unbound mass to about $\sim 65 \%$
\cite{rosswog08a}. Before the energy contained in the bound matter can be released, an accretion
disk has to form. On returning to the black hole, the orbits of the bound matter become focussed
once more towards peri-center. The spread of specific energies across the accretion stream
is very large \cite{rees88} so that matter that passes on the inside of the stream, i.e. closer to the black
hole, is considerably stronger bound and subsequently only reaches moderate apocentre distances 
while matter on the outside track is relaunched into a substantially wider orbit. Thus, after having passed
the hole for the second time, the matter stream spreads in a fan-like manner \cite{rees88,rosswog08d}.
It subsequently collides with the still infalling material and this self-interaction produces an angular
momentum redistribution shock that circularizes the accretion flow, see e.g. Fig. 4 in 
\cite{rosswog08b}. In the early stages, mass is fed towards the black hole at a rate that carries the imprint
of the internal structure of the star \cite{lodato08,ramirezruiz08}, at later stages the rate falls off $\propto t^{-5/3}$ 
\cite{rees88,phinney89}. Once a disk has formed, it evolves under the influence of viscosity, radiatively driven 
winds and the fallback rate by which it is fed. We expect a luminosity that is comparable to the Eddington 
value, $L_{\rm Edd} \simeq 10^{41}{\rm erg/s} \; (M_{\rm bh}/1000 M_\sun)$, typically lasting for a few months.

\section{Summary and discussion}
We have explored in detail the fate of white dwarfs that approach a moderately-massive black hole close enough 
to become tidally disrupted. Above a limiting mass of $\sim 10^5$ \msun, the detailed value depending on 
the white dwarf, the black hole swallows the star as a whole before disrupting it. Thus 
white dwarfs represent a precious tool to probe the existence of moderate-mass black holes. White dwarfs 
that penetrate the tidal radius deeply enough are heated by the tidal compression and ensuing shock to the 
temperatures of nuclear statistical equilibrium. In favorable cases more than the gravitational self-binding 
energy of the star can be released via nuclear reactions thus triggering a thermonuclear explosion of the 
white dwarf. The amount of iron-goup nuclei produced is rather sensitive to the densities at which nuclear 
burning occurs, and therefore to the white dwarf mass. The amount produced in the exploding cases of our 
simulation set \cite{rosswog08d} ranges from 0.03 \Msun for a 0.2 \Msun pure helium white dwarf to 0.66 
\Msun for a 1.2 \Msun carbon-oxygen white dwarf, i.e. if composed of mainly $^{56}$Ni the favorable cases 
produce amounts comparable to standard type Ia supernovae \cite{mazzali07}.\\
These tidally-induced supernovae are, however, in several respects different from what is considered
a "normal" type Ia supernova. First, they are not restricted to progenitor masses close to the Chandrasekhar limit.
We find exploding examples throughout the full explored mass range from 0.2  to 1.2 \msun. Typical examples 
of these explosions are close to the peak of the white dwarf mass distribution, i.e. near 0.6 \msun, or slightly above 
this value, since in the dense stellar environments around such black holes close encounters between stars may
lead to mass segregation \cite{binney08}. Thus the average progenitor will be less massive than for a normal type Ia. 
Second, and closely related to the first point, the progenitor composition is not restricted to carbon-oxygen.
Also pure helium white dwarfs can explode via this mechanism, but since white dwarfs near and slightly beyond the
peak of the mass distribution are thought to consist of carbon and oxygen this is the most common progenitor
composition. Due to the initial chemical layering of the white dwarfs,  the outer shells of the most common
explosion remnants consist of unburned, highly carbon-enriched ($X_C \sim 0.8$) material. Third, the 
remnant geometry is most likely much less spherical than a standard type Ia, therefore the optical lightcurve 
should be rather unique as a result of the radiating material being highly squeezed into the orbital plane. The 
large velocities of the ejected debris ($> 10^4$ km/s) should produce large Doppler shifts. And last but not least, 
this peculiar type of thermonuclear supernova is accompanied by an X-ray flare close to the Eddington luminosity 
that lasts for a few months.\\
The estimated event rates for this type of transient are  $\sim 10^{-3}$ of the type Ia supernova rate. Though 
being substantially less frequent, they occur often enough to warrant a search for this new class of optical 
transient. Upcoming supernova searches hope to discover several thousand to several hundred thousand type 
Ia-like events \cite{riess06,aldering07} per year. Chances are promising to find examples of tidally-induced thermonuclear 
supernovae among them. Maybe, the recently detected carbon-rich transient SCP 06F6 accompanied by an X-ray 
signal \cite{gaensicke08} is already the first example for this class of object.\\

{\bf Acknowledgements} We thank Holger Baumgardt, Peter Goldreich, Jim Gunn,
Piet Hut, Dan Kasen, Bronson Messer and Martin Rees for very useful
discussions. E. R. acknowledges support from the DOE Program for
Scientific Discovery through Advanced Computing (SciDAC;
DE-FC02-01ER41176). The simulations presented in this paper were 
performed on the JUMP computer of the H\"ochstleistungsrechenzentrum J\"ulich.
Oak Ridge National Laboratory is managed by UT-Battelle, LLC, for the  
U.S. Department of Energy under contract DE-AC05-00OR22725.\\

{\bf References}
\bibliography{astro_SKR}
\bibliographystyle{elsart-num.bst}
\end{document}